\documentclass[aps,pra,twocolumn,letterpaper,superscriptaddress,10pt]{revtex4-2}
\usepackage{amssymb,amsthm,amsmath,amsfonts}
\usepackage{graphicx,ulem,enumerate,bbm,bm,mathptmx}
\usepackage[pdftex,dvipsnames,usenames]{xcolor}
\usepackage[colorlinks=true,urlcolor=blue,citecolor=blue,linkcolor=blue]{hyperref}
\usepackage{dsfont}

\begin{document}

\title{%
    Quantum sampling in hybrid light-matter systems with mixed statistics
}

\author{Franziska Barkhausen}
    \email{fbarkhau@mail.uni-paderborn.de}
    \affiliation{Institute for Photonic Quantum Systems (PhoQS), Department of Physics and Center for Optoelectronics and Photonics Paderborn (CeOPP), Paderborn University, Warburger Strasse 100, D-33098 Paderborn, Germany}

\author{Laura Ares}
    \email{laura.ares.santos@uni-paderborn.de}
    \affiliation{Institute for Photonic Quantum Systems (PhoQS), Department of Physics and Center for Optoelectronics and Photonics Paderborn (CeOPP), Paderborn University, Warburger Strasse 100, D-33098 Paderborn, Germany}

\author{Stefan Schumacher}
    \affiliation{Institute for Photonic Quantum Systems (PhoQS), Department of Physics and Center for Optoelectronics and Photonics Paderborn (CeOPP), Paderborn University, Warburger Strasse 100, D-33098 Paderborn, Germany}
    \affiliation{Wyant College of Optical Sciences, University of Arizona, Tucson, Arizona 85721, USA}

\author{Jan Sperling}
    \affiliation{Institute for Photonic Quantum Systems (PhoQS), Department of Physics and Center for Optoelectronics and Photonics Paderborn (CeOPP), Paderborn University, Warburger Strasse 100, D-33098 Paderborn, Germany}

\date{\today}

\begin{abstract}
    Boson sampling is one of the most prominent methods to verify quantum advantage, harnessing the computational complexity of the permanent of a network matrix.
    Similar quantum sampling problems utilize other expressions for the output probabilities, such as the fermion-based analog applying determinants.
    In this work, we study sampling setups of interacting light-matter systems, combining fermion and boson sampling.
    The mixed network input consists of fermionic and bosonic excitations.
    Sampling the output probability of the modes then yields contributions that are neither purely determinant nor permanent but general immanants.
    Specifically, we prove that these immanant contributions are linearly independent from determinants and permanents, rendering our mixed sampling network an ideal candidate for light-matter quantum application beyond pure fermion and boson sampling.
    Furthermore, quantum sampling with statistics that are neither bosonic nor fermionic results in expressions which even exceed immanants, further extending the functionality of quantum samplers.
\end{abstract}

\maketitle


\section{Introduction}
\label{sec:Introduction}

    Boson sampling \cite{SAAA13,GD15} has become one of the well-established benchmarks that demonstrates the advantage of quantum computers over classical algorithms by building a quantum setup that is hard to simulate using classical computers.
    Boson sampling is based on calculating the permanent of a large unitary matrix, and thus of the underlying boson sampling network \cite{MRAZ94}, which is $\#\mathsf{P}$ hard to compute for classical algorithms, requiring an exponential overhead in computational time, under the assumption that the polynomial hierarchy does not collapse, $\mathsf{P}\neq\mathsf{NP}$.
    Early works showed the inherent complexity when determining the permanent \cite{PB13,LV79}, and continued effort is dedicated to propose optimized algorithms for its computation \cite{PLKM22}.

    The fermionic analog of boson sampling is fermion sampling \cite{ODMZ22}, for which the determinant is used.
    The complexity of the determinant is polynomially bounded and thus easier to compute classically than the permanent.
    The encompassing expression for both determinants and permanents, the immanant of a matrix \cite{DLAR34}, can be used in quantum interferometry for determining coincidence probabilities \cite{KGS20,TGGS13}, extending to other applications, too \cite{GGC03}.
    The complexity of the immanant \cite{PB00,SMCM13,NRA13} generally depends on whether it is more similar to the determinant or permanent \cite{SMCM13,WH85}.

    Even in the purely bosonic context, there exists a variety of different boson sampling approaches, such as scattershot boson sampling \cite{LLRRBR14}, Gaussian boson sampling \cite{HKS17,LLRROR14}, and non-Gaussian boson sampling \cite{CHIJ24}, which can be also unified \cite{BMABS25}.
    Quantum sampling algorithms are not limited to photons, e.g., evident through atomic boson sampling \cite{YGE24}.
    Moreover, a sampling setup does not have to consist of a single type of quantum field.
    For example, in Ref. \cite{VK24}, the concept of coupled bosonic particles is put forward, merging photons and condensed atoms.
    There are also generalized approaches jointly applicable to bosons and fermions, e.g., in Ref. \cite{SNSG22}, where expressions combine determinants and permanents as contributions to coincidence rates from an interferometer.
    Approaches beyond fermions and bosons are also found in Refs. \cite{TM17,ZWKH25}, where non-trivial exchange symmetries and parastatistics are exploited.

    In quantum physics, combined light-matter systems are paramount for studying interactions in heterogeneous systems and for establishing quantum interfaces \cite{LB23}.
    For example, the pioneering Jaynes--Cummings model \cite{JC63} describes the most fundamental exchange of excitations between bosons, i.e., one optical mode, and fermions, i.e., one electronic mode.
    The eigenstates of this model are known as dressed states \cite{VW06}.
    In general, joint excitations of such hybrid systems result in the notion of quasiparticles and collective excitations in condensed matter physics, such as excitons and polaritons, magnons, phonons, plasmons, etc \cite{LP21}.
    The light-matter hybrid nature can exhibit interesting entanglement properties \cite{BASS25} and allows for complex forms of entanglement when propagated in large-scale networks \cite{FAS25}.
    A relatively recent study \cite{jabbour23} focused on the interplay of mixed fermion-boson transition probabilities in a linear interferometer, finding remarkable relations between the determinant and permanent.

    In this work, we study general quantum sampling problems to advance applications of light-matter systems, producing expressions, including immanants, beyond determinants and permanents.
    To this end, we combine heterogeneous inputs with different exchange symmetries via a linear network, detecting both fermions and bosons at the output.
    In such composite sampling setups, we quantify the non-permanent and non-determinant character of the obtained immanant variants.
    We explore cases of such mixed fermion--boson systems in which the resulting immanants cannot be written as linear combinations of determinants and permanents.
    Generalizations to other exchange symmetries are discussed, specifically quasiparticle descriptions that are shown to exceed even immanant sampling.
    Thereby, heterogeneous networks are shown to offer a rich physical platform for quantum sampling algorithms, using light-matter interactions to benchmark interfaces in quantum science and technology.

    The paper is structures as follows:
    First, we introduce the linear network system with fermionic and bosonic excitations operators in Sec. \ref{sec:System}.
    General expression for the output probabilities for genral anyonic exchange symmetries are derived in Sec. \ref{sec:Generalization}.
    In Sec. \ref{sec:Results}, concrete examples are given for a specific number of input modes and particles and its corresponding output, and we verify the linear independence of our expressions from permanents and determinants as well as their basis properties.
    Section \ref{sec:Quasiparticles} is dedicated to quantum sampling from quasiparticle descriptions, which introduces expressions more powerful than obtained from immanants.
    Eventually, we conclude in Sec. \ref{sec:Conclusion}.


\section{Hybrid system under study}
\label{sec:System}

\paragraph{Modes and transformations.}

    Suppose we have a linear network with $F$ fermionic and $B$ bosonic input and output modes.
    Through this network, purely bosonic and purely fermionic modes are superimposed, producing hybrid output modes.
    The fermionic and bosonic particles at the inputs of a mode are described using the creation operators $\hat{f}_j$ and $\hat{b}_i$, with $j\in \{1,\ldots,F\}$ and $i\in \{1,\ldots,B\}$, respectively.
    The fermionic and bosonic field operators obey the fundamental commutator ($[\,\cdot\,,\,\cdot\,]$) and anti-commutator ($\{\,\cdot\,,\,\cdot\,\}$) relations
    \begin{equation}
        \label{eq:CommutationRelations}
    \begin{aligned}
    	\{\hat f_i,\hat f_{i'}\}=0=\{\hat f_i^\dag,\hat f_{i'}^\dag\},
    	\quad
    	[\hat b_j,\hat b_{j'}]=0=[\hat b_j^\dag,\hat b_{j'}^\dag],
    	\\
    	\{\hat f_i,\hat f_{i'}^\dag\}=\delta_{i,i'},
    	\quad\text{and}\quad
    	[\hat b_j,\hat b_{j'}^\dag]=\delta_{j,j'},
    \end{aligned}
    \end{equation}
    where $\delta$ denotes the Kronecker symbol.
    The unitary network performs a linear transformation of
    \begin{equation}
        \label{eq:unitaryfb}
        \begin{bmatrix}
    		\hat f_1^{\dag}
    		\\
            \vdots
            \\
            \hat f_{F}^{\dag}
            \\
    		\hat b_1^{\dag}
    		\\
            \vdots
            \\
            \hat b_{B}^{\dag}
    	\end{bmatrix}
        \mapsto
        \begin{bmatrix}
            \phi_{1,1} & \cdots & \phi_{1,F}
            &
            \gamma_{1,1} & \cdots & \gamma_{1,B}
            \\
            \vdots & \ddots & \vdots
            &
            \vdots & \ddots & \vdots
            \\
            \phi_{F,1} & \cdots & \phi_{F,F}
            &
            \gamma_{F,1} & \cdots & \gamma_{F,B}
            \\
            \tilde{\gamma}_{1,1} & \cdots & \tilde{\gamma}_{1,F}
            &
            \beta_{1,1} & \cdots & \beta_{1,B}
            \\
            \vdots & \ddots & \vdots
            &
            \vdots & \ddots & \vdots
            \\
            \tilde{\gamma}_{B,1} & \cdots & \tilde{\gamma}_{B,F}
            &
            \beta_{B,1} & \cdots & \beta_{B,B}
        \end{bmatrix}
        \begin{bmatrix}
            \hat f_1^{\dag}
            \\
            \vdots
            \\
            \hat f_{F}^{\dag}
            \\
            \hat b_1^{\dag}
            \\
            \vdots
            \\
            \hat b_{B}^{\dag}
        \end{bmatrix}.
    \end{equation}
    Therein, the linear map $U\in\mathbb C^{M\times M}$, with $M=F+B$, is described by blocks of entries, with $\phi_{i,i'}$ for fermion-fermion interactions and $\beta_{j,j'}$ for boson-boson interactions as well as $\gamma_{i,j}$ and $\tilde{\gamma}_{j,i}$ to account for interactions between the fermionic and bosonic field operators.

\paragraph{Input states and measurement projections.}

    The vacuum state is denoted by $|\mathrm{vac}\rangle$, and it remains unchanged under the linear map under consideration.
    The input states $|\Psi\rangle$ under study are excitations of the individual modes, such as the number state
    \begin{equation}
    \begin{aligned}
        \label{eq:InputType}
        |\Psi\rangle
        ={}&
        \hat f_{m}^\dag \cdots \hat f_{1}^\dag
        \hat{b}^{\dag}_n\dots \hat{b}^{\dag}_1|\mathrm{vac}\rangle
        \\
        ={}&
        |\underbrace{1,\dots,1}_{m},\underbrace{0,\dots,0}_{F-m}\rangle \otimes |\underbrace{1,\dots,1}_{n},\underbrace{0,\dots,0}_{B-n}\rangle,
    \end{aligned}
    \end{equation}
    with $m$ fermions, $n$ bosons, and all other $(F-m)+(B-n)$ modes remaining empty.
    For the input in Eq. \eqref{eq:InputType}, the output state takes the form
    \begin{equation}
    \begin{aligned}
        \label{eq:PsioutFB}
        |\Psi\rangle
        \mapsto{}&
        \prod_{k=1}^m\left(
            \sum_{i=1}^F\phi_{k,i}\hat f_i^\dag
            +
            \sum_{j=1}^B\gamma_{k,j}\hat b_j^\dag
        \right)
        \\
        {}&\times
        \prod_{l=1}^n\left(
            \sum_{i=1}^F\tilde{\gamma}_{k,i}\hat f_i^\dag
            +
            \sum_{j=1}^B\beta_{k,j}\hat b_j^\dag
        \right)
        |\mathrm{vac}\rangle
        =|\Psi'\rangle.
    \end{aligned}
    \end{equation}

    On the measurement side, we consider detections of single excitations of the output modes, such as given through projections of the output state with $\langle\Pi|=\langle\mathrm{vac}|\hat f_1\cdots\hat f_{m'}\hat b_1\cdots\hat b_{n'}$.
    Note that we restrict ourselves to at most a single excitation per output mode.
    Thus, Born's rule tells us that the output probability is given by
    \begin{equation}
        \label{eq:BornRule}
        p=\left|\langle \Pi|\Psi'\rangle\right|^2,
    \end{equation}
    which is zero when the number of input excitations, $m+n$, does not match the number of detected excitations, $m'+n'$.
    We focus on cases where the number of particle measured coincides with number of incident particles, $N=m+n=m'+n'$.

\paragraph{Immanants.}

    The variety of boson sampling approaches offers different tools for analyzing the measured output probability.
    For example, Gaussian boson sampling results in output probability distribution determined through Hafnians \cite{GIQ19,HKS17}, and fermions would similarly apply Pfaffnians.
    For single excitations, as studied here, common boson and fermion sampling schemes are based on the permanent and determinant, respectively.
    As mentioned in the introduction, generalized interferometer coincidence of fermion and boson rates can apply immanants \cite{SNSG22}, the generalization of determinants and permanents.
    A general immanant of a matrix $A\in\mathbb C^{N\times N}$ is given by
    \begin{equation}
        \label{eq:immmfunction}
        \mathrm{imm}(A)=\sum_{\sigma\in S_N}\chi(\sigma)\prod_{i=1}^N A_{i,\sigma(i)},
    \end{equation}
    with the group $S_N$ of permutation of $N$ elements and a function $\chi$ mapping permutations to generally complex numbers.
    Specifically, for the constant function, $\chi(\sigma)=1$, we obtain the permanent, $\mathrm{imm}(A)=\mathrm{perm}(A)$, and we obtain the determinant, $\mathrm{imm}(A)=\det(A)$, when $\chi(\sigma)=\mathrm{sgn}(\sigma)$ is the sign function, which is one for an even permutation and negative one for an odd permutation.
    Since Eq. \eqref{eq:PsioutFB} includes fermionic and bosonic contributions as well as combinations thereof, the output state $|\Psi'\rangle$ contains mixed excitation and requires generalized expressions of immanants in the probability $p$.

    In general, the immanant in Eq. \eqref{eq:immmfunction} is an $N$-linear function from matrices to complex numbers \cite{KRGS11}.
    The weight of the contribution of each permutation $\sigma$ is described by the function $\chi(\sigma)$.
    In some definitions of immanants, this function $\chi=\chi_\lambda$ is a so-called irreducible character of the corresponding symmetric group $S_N$ for the partition $\lambda=(\lambda_1,\dots,\lambda_L)$ of length $L$ of $N$ numbers and was originally introduced in Ref. \cite{DLAR34}.
    The number of immanants of a matrix $A\in \mathbb{C}^{N\times N}$ is equal to the number of partitions of $N$.
    As mentioned above, the important special case of the permanent is obtained for the trivial irreducible character $\chi_{\lambda}=1$, each permutation is multiplied by $1$, and the determinant is obtained when using the parity of the permutation, i.e., $\chi_{\lambda}=\mathrm{sgn}$.
    A common example of a immanant distinct from the determinant and permanent for $N=3$ is
    \begin{equation}
        \label{eq:Std3Imm}
        \mathrm{imm}(A)
        =
        2A_{1,1}A_{2,2}A_{3,3}
        -A_{1,2}A_{2,3}A_{3,1}
        -A_{1,3}A_{2,1}A_{3,2},
    \end{equation}
    where $\chi_\lambda$ assigns the value of two, zero, and negative one for cycles of lengths one, two, and three, respectively, that decompose a permutation.
    See also Table \ref{tab:ThreeDimPermutations} for the case $N=3$.

\begin{table}
    \caption{%
        For $N=3$, the permutations $\sigma\in S_N$ and their characteristics are shown.
        The first column shows the one-line notation, listing the outcomes $\sigma(j)$ in order from $j=1$ to $j=N$.
        The cycle notation is provided in the second column, where, for example, $(132)$ describes a mapping $1\stackrel{\sigma}{\mapsto}3\stackrel{\sigma}{\mapsto}2\stackrel{\sigma}{\mapsto}1$.
        The last three columns provide the function $\chi=\chi_\lambda$ for the permanent, determinant, and Eq. \eqref{eq:Std3Imm}.
    }\label{tab:ThreeDimPermutations}
    \begin{tabular}{ccccc}
        \hline \hline
        one-line notation & cycle notation & $\mathrm{\chi_\lambda}=1$ & $\mathrm{\chi_\lambda}=\mathrm{sgn}$ & $\chi_\lambda$ for Eq. \eqref{eq:Std3Imm}
        \\
        \hline
        $[123]$ & $(1)(2)(3)$ & $1$ & $1$ & $2$
        \\
        $[213]$ & $(12)(2)$ & $1$ & $-1$ & $0$
        \\
        $[321]$ & $(13)(2)$ & $1$ & $-1$ & $0$
        \\
        $[132]$ & $(1)(23)$ & $1$ & $-1$ & $0$
        \\
        $[231]$ & $(123)$ & $1$ & $1$ & $-1$
        \\
        $[312]$ & $(132)$ & $1$ & $1$ & $-1$
        \\
        \hline\hline
    \end{tabular}
\end{table}

    Immanants describe a broad variety of related structures, inlcuding the families of Temperley--Lieb immannants with totally nonnegative functions \cite{BRMS05} and Kazhdan--Lusztig immanants \cite{BRMS06}.
    One expample of Temperley-Lieb immanants is the determinant \cite{BRMS05}.
    Other types of immanant structures include immanant varieties \cite{DBPS24} and quasi-immanants \cite{JMC25}.
    In this work, we apply immanant functions, determined by $\chi$ in Eq. \eqref{eq:immmfunction}, beyond irreducible characters $\chi_\lambda$.
    The main difference is that the irreducible characters have the same value for cycles of equal length $L$.
    These polynomial functions are have been widely studied in the literature \cite{MMHM65,RC21,LW23}.
    In contrast, we allow for non-trivial functions that are linearly independent from such immanants based on irreducible characters.
   Note that, in Sec. \ref{sec:Quasiparticles}, we even find physical systems in which the expression in Eq. \eqref{eq:immmfunction} has to be generalized as the resulting probabilities include quasiparticle contributions that exceed permutations of matrix entries.


\section{Generalized approach}
\label{sec:Generalization}

    In the framework discussed so far, we established the composite fermion boson network explicitly using separate excitation operators with the corresponding exchange symmetries.
    In this section, we generalize this approach, even including anyons, for finding the probabilities that generally determine the quantum sampling probabilities.

\paragraph{Field operators and network unitary.}

    We can introduce general creation operators $\hat{a}_j^\dag$ for the field excitations, with $j\in\{1,\ldots,M\}$ indicating the modes, without explicitly distinguishing between fermions, bosons, or otherwise.
    The (anti-)commutator relations in Eq. \eqref{eq:CommutationRelations} can be generalized as follows:
    \begin{align}
        \label{eq:conv}
        \hat{a}^\dag_i\hat{a}^\dag_j
        ={}&
        s_{i,j}\hat{a}^\dag_j\hat{a}^\dag_i,
        \\
        \label{eq:DefinitionS}
        \text{where }
        s_{i,j}
        ={}&
        \left\lbrace
        \begin{array}{ll}
            -1 & \text{ for both $\hat a_i^\dag$ and $\hat a_j^\dag$ fermionic},
            \\
            1 & \text{ otherwise},
        \end{array}
        \right.
    \end{align}
    with $s_{i,j}=1$ accounting for both modes being bosonic as well as modes from commuting fields, such as one bosonic mode $i$ and one fermionic mode $j$.
    The elements of $s$ can be extended with components akin to $s_{i,j}=e^{-i\theta}$ ($0<\theta<\pi$) for two modes of anyons \cite{LM77,W82}, $\hat a_i^\dag$ and $\hat a_j^\dag$, rendering this approach rather flexible.
    It is also worth mentioning that the number $M=F+B$ is the total of the bosonic and fermionic modes---plus the number of anyonic modes if needed.

    Equation \eqref{eq:unitaryfb}, describing the network, can be generalized in a similar manner.
    That is, the unitary matrix $U \in \mathbb{C}^{M\times M}$ is the input-output map for the network under study, consisting of $M=F+B$ channels and resulting in the evolution
    \begin{equation}
        \label{eq:GeneralUnitary}
        \hat{a}_j^\dag\mapsto \sum^{M}_{k=1}U_{j,k} \hat{a}_k^\dag,
    \end{equation}
    where we generally not distinguish between bosons, fermions, and anyons.
    The expression in Eq. \eqref{eq:GeneralUnitary} is going to result in the explicit expressions for immanants that can be obtained from light-matter interactions.

\paragraph{Input and output states.}

    Again, we take single excitations of individual modes as the input state.
    Let $j_1<\cdots<j_N$ determine which modes are initially excited.
    Then, we have
    \begin{equation}
        |\Psi\rangle
        =
        \hat a^\dag_{j_N}\cdots \hat a^\dag_{j_1}
        |\mathrm{vac}\rangle
    \end{equation}
    as the $N$-particle input, where we assume $N\leq M$.
    The output state $|\Psi'\rangle$ then formally reads as
    \begin{equation}
         |\Psi'\rangle
         =
         \prod^N_{n=1}\left(
            \sum_{k=1}^M U_{j_n,k}\hat a_k^\dag
         \right)|\mathrm{vac}\rangle,
    \end{equation}
    by applying Eq. \eqref{eq:GeneralUnitary}.

    To evaluate the output expression further, we consider the operator part alone.
    We compute
    \begin{equation}
        \label{eq:ManipulationOutputOne}
    \begin{aligned}
        {}&\prod^N_{n=1}\left(
            \sum_{k}U_{j_n,k}\hat a_k^\dag
        \right)
        =
        \sum_{k_1,\ldots,k_N}
        \left(\prod_{n=1}^N U_{j_n,k_n}\right)
        \hat a^\dag_{k_N}\cdots\hat a^\dag_{k_1}
        \\
        ={}&
        \sum_{\sigma\in S_N}\sum_{\substack{
            k_1,\ldots,k_N:
            \\ k_1<\cdots<k_N
        }}
        \left(\prod_{n=1}^N U_{j_n,\,k_{\sigma(n)}}\right)
        \hat a^\dag_{k_{\sigma(N)}}\cdots\hat a^\dag_{k_{\sigma(1)}}
        \\
        {}&+(\text{higher-excitation terms}).
    \end{aligned}
    \end{equation}
    Therein, the first contribution accounts for all excitations in which the indices $k_n$ are all different, which can be achieved by considering the order $k_1<\cdots<k_N$ and all permutations $\sigma$ thereof;
    the higher-excitation terms include at least two elements $k_n=k_{n'}$ that are identical.
    For fermions, Pauli exclusion yields $\hat a_{k_n}^{\dag\,2}=0$, disallowing for identical states and removing such higher-excitation terms.
    Otherwise, we can consider networks which dilute the excitations, $M\gg N$, such that it becomes unlikely that two particles end up in the same output mode, and the measurement projection $\langle\Pi|$ is based on single excitation, post-selecting for events that do not include higher excitations.
    We can further look at the operator term in Eq. \eqref{eq:ManipulationOutputOne} and apply induction to order the operators via the general exchange symmetry from Eq. \eqref{eq:conv}, which allows us to write
    \begin{equation}
        \label{eq:ManipulationOutputTwo}
        \hat a^\dag_{k_{\sigma(N)}}\cdots\hat a^\dag_{k_{\sigma(1)}}
        =
        \underbrace{
            \left(\prod_{n=1}^N\prod_{\substack{
                m\in\{1,\ldots,M\}:
                \\
                \sigma(m)>n
            }}s_{k_n,\,k_{\sigma(m)}}\right)
        }_{\stackrel{\text{def.}}{=}\chi(\sigma)}
        \hat a_{k_N}^\dag\cdots\hat a_{k_1}^\dag,
    \end{equation}
    defining the function $\chi$ needed for the immanant in Eq. \eqref{eq:immmfunction}.

    Combining all above calculations, we obtain the output state
    \begin{align}
        \nonumber
        |\Psi'\rangle
        ={}&
        \sum_{\substack{
            k_1,\ldots,k_N:
            \\ k_1<\cdots<k_N
        }}
        \left(\sum_{\sigma\in S_N}\chi(\sigma)
        \prod_{n=1}^N U_{j_n,\,k_{\sigma(n)}}\right)
        \hat a_{k_N}^\dag\cdots\hat a_{k_1}^\dag|\mathrm{vac}\rangle
        \\
        {}&+(\text{higher-excitation terms}).
    \end{align}
    For each choice $k_1<\cdots<k_N$ and initial excitations $j_1<\cdots<j_N$, the expression in the parenthesis corresponds to the immanant  in Eq. \eqref{eq:immmfunction}, using the function $\chi$ as defined in Eq. \eqref{eq:ManipulationOutputTwo} and the $N\times N$ matrix $A$ with the entries
    \begin{equation}
        A_{n,m}=U_{j_{n},k_{m}},
        \text{ for }
        n,m\in\{1,\ldots,N\},
    \end{equation}
    which is a submatrix of $U\in\mathbb C^{M\times M}$ using the rows of the initial exitations and columns of the output.
    For a measurement of $\langle\Pi|=\langle\mathrm{vac}|\hat a_{k_1}\cdots\hat a_{k_N}$, we thus find
    \begin{equation}
        \label{eq:GeneralProbability}
        p=\left|\langle\Pi|\Psi\rangle\right|^2
        =\left|\mathrm{imm}(A)\right|^2,
    \end{equation}
    which combines fermions and bosons, as well as anyons if one wishes.
    Again, the explicit form of $\chi$ that defines the immanant of arbitrary interacting modes is provided in Eq. \eqref{eq:ManipulationOutputTwo} and shall be explored with concrete examples in the following section.
    See also Ref. \cite{SNSG22} in this context, where Fourier methods and Young tableaux were used to determine coincidence rates from heterogeneous interferometers.

\paragraph{Preliminary summary.}

    From our derivation, when restricting ourselves to $M=F+B$ modes of fermions ($F$) and bosons ($B$), we can identify three types of general matrix functions in the output distribution $p$.
    On the one hand, we find determinants ($F=M$ and $B=0$) and permanents ($F=0$ and $B=M$).
    On the other hand, there can be mixed contributions that cannot be classified as a determinant or permanent because some $s_{j,k}$ are positive and some are negative.
    This mixing is determined by the input particles ($j_1,\ldots, j_N$) as well as the measurement projection ($k_1,\ldots,k_N$).
    In general, the permutations induced by the relations given by $s_{j,k}$ in Eq. \eqref{eq:DefinitionS} create non-trivial output probabilities.
    Thus, the mixed light--matter system with interactions requires and results in additional matrix immanants.
    These matrix functions are not necessarily based on the irreducible characters ($\chi\neq\chi_\lambda$)---as seen in the following and reflecting the complexity and richness of such hybrid light-matter quantum systems.


\section{Applications}
\label{sec:Results}

    In this section, we study and quantify to which extend quantum sampling problems in hybrid light-matter systems are distinct from the original sampling in purely bosonic and fermionic scenarios.
    Specifically, we analyze the behavior of interferometers based on the function $\chi$, which we are going interpret as a vector, to characterize the kind of immanant, Eq. \eqref{eq:immmfunction}, that we find in specific scenarios.

\subsection{Three modes with three particles}

    Firstly, we study a sort of minimal example with three modes of fermionic and bosonic character, i.e., $M=3$.
    In addition, we consider $N=3$ excitations, each in one of the six modes, which vary from purely fermionic and bosonic to the mixed cases, being the more relevant scenario for us.

    The general unitary network from Eqs. \eqref{eq:unitaryfb} and \eqref{eq:GeneralUnitary} reads
    \begin{equation} 
    \label{eq:exampleumap}
        \begin{bmatrix}
    		\hat a_1^{\dag}
    		\\
            \hat a_2^{\dag}
    		\\
            \hat a_3^{\dag}
    	\end{bmatrix}
        \mapsto
        \begin{bmatrix}
            U_{1,1} & U_{1,2} & U_{1,3}
            \\
            U_{2,1} & U_{2,2} & U_{2,3}
            \\
            U_{3,1} & U_{3,2} & U_{3,3}
        \end{bmatrix}
        \begin{bmatrix}
    		\hat a_1^{\dag}
    		\\
            \hat a_2^{\dag}
    		\\
            \hat a_3^{\dag}
    	\end{bmatrix}
        .
    \end{equation}
    Having three excitations ($j_1=1$, $j_2=2$, $j_3=3$) at the input, according to the previous derivation in Eqs. \eqref{eq:ManipulationOutputOne} and \eqref{eq:ManipulationOutputTwo}, then can be recast into the output form ($k_1=1$, $k_2=2$, $k_3=3$)
    \begin{equation}
        \hat a_{3}^\dag\hat a_{2}^\dag\hat a_{1}^\dag
        \mapsto
        \mathrm{imm}(U)
        \hat a_{3}^\dag\hat a_{2}^\dag\hat a_{1}^\dag
        +(\text{higher-excitation terms}),
    \end{equation}
    using the immanant that accounts for the terms for all permutations in $S_3$ (cf. Table \ref{tab:ThreeDimPermutations}), with
    \begin{equation}
    \begin{aligned}
        \mathrm{imm}(U)
        ={}&
        U_{1,1}U_{2,2}U_{3,3} 
        \\
        {}&+
        U_{1,2}U_{2,1}U_{3,3} \, s_{1,2} 
        \\
        {}&+
        U_{1,3}U_{2,2}U_{3,1} \, s_{1,2}s_{1,3}s_{2,3} 
        \\
        {}&+
        U_{1,1}U_{2,3}U_{3,2} \, s_{2,3} 
        \\
        {}&+
        U_{1,2}U_{2,3}U_{3,1} \, s_{1,2}s_{1,3} 
        \\
        {}&+
        U_{1,3}U_{2,1}U_{3,2} \, s_{1,3}s_{2,3} 
        .
    \end{aligned}
    \end{equation}
    In addition, it is convenient to introduce a six-dimensional vector with entries $\chi(\sigma)$ for permutations $\sigma\in S_3$ as listed in Table \ref{tab:ThreeDimPermutations}.
    This vector reads
    \begin{equation}
        \vec \chi
        =
        \begin{bmatrix}
            1 
            \\
            s_{1,2} 
            \\
            s_{1,2}s_{1,3}s_{2,3} 
            \\
            s_{2,3} 
            \\
            s_{1,2}s_{1,3} 
            \\
            s_{1,3}s_{2,3} 
        \end{bmatrix}.
    \end{equation}

    For all cases discussed in the following, we consider
    \begin{equation}
        |\Psi\rangle=|1,1,1\rangle
        \text{ and }
        \langle\Pi|=\langle 1,1,1|
    \end{equation}
    for the input state and the output measurement, respectively.
    Propagating the state as described above and using Born's rule, Eq. \eqref{eq:BornRule}, we find the detection probability via Eq. \eqref{eq:GeneralProbability}.

    Next, we can consider the different scenarios.
    For measurements of three fermions, we have $s_{1,2}=s_{1,3}=s_{2,3}=-1$ and
    \begin{equation}
        \vec \chi
        =
        \begin{bmatrix}
            +1 & -1 & -1 & -1 & +1 & +1
        \end{bmatrix}^\mathrm{T}=\vec\chi_{\det}.
    \end{equation}
    Using Table \ref{tab:ThreeDimPermutations} and Eq. \eqref{eq:GeneralProbability}, we see that this scenario corresponds to $p=|\det(U)|^2$, as one would expect.
    Next, we consider two fermions and one boson, resulting in an anti-commuting $s_{1,2}=-1$ and commuting $s_{1,3}=s_{2,3}=1$, thus
    \begin{equation}
        \label{eq:ImmanantExample1}
        \vec \chi
        =
        \begin{bmatrix}
            +1 & -1 & -1 & +1 & -1 & +1
        \end{bmatrix}^\mathrm{T}=\vec\chi_\mathrm{FFB},
    \end{equation}
    which is neither a permanent, determinant, nor the common immanant in Eq. \eqref{eq:Std3Imm}; compare also with Table \ref{tab:ThreeDimPermutations}.
    Lastly, when we consider two bosons and one fermion or three bosons, we find all modes commuting, $s_{1,2}=s_{1,3}=s_{2,3}=1$, resulting in the permanent $p=|\mathrm{perm}(U)|^2$, likewise
    \begin{equation}
        \vec \chi
        =
        \begin{bmatrix}
            +1 & +1 & +1 & +1 & +1 & +1
        \end{bmatrix}^\mathrm{T}=\vec\chi_\mathrm{perm}.
    \end{equation}

    In addition to the determinant and permanent, we find a non-trivial immanant for two fermions and one boson, which is not even of the character given in Eq. \eqref{eq:Std3Imm}.
    Even further, we can decompose the immanant given through Eq. \eqref{eq:ImmanantExample1} as
    \begin{equation}
        \label{eq:ThreeParticleDecomposition}
    \begin{aligned}
        \vec \chi_\mathrm{FFB}
        =\frac{2}{6}\vec\chi_{\det}
        +\frac{0}{6}\vec\chi_\mathrm{perm}
        {}&+\frac{2}{6}\vec\chi_\text{Eq. \eqref{eq:Std3Imm}}
        +\vec\chi_\mathrm{residuum},
        \\
        \text{where }
        \vec\chi_\text{Eq. \eqref{eq:Std3Imm}}={}&
        \begin{bmatrix}
            2 & 0 & 0 & 0 & -1 & -1
        \end{bmatrix}^\mathrm{T}
        \\
        \text{and }
        \vec\chi_\mathrm{residuum}={}&-\frac{1}{3}
        \begin{bmatrix}
            0 & 2 & 2 & -4 & 3 & 3
        \end{bmatrix}^\mathrm{T},
    \end{aligned}
    \end{equation}
    with the residuum being the linearly independent---in fact, orthogonal---part of the two-fermion-one-boson immanant determined through $\vec \chi_\mathrm{FFB}$.
    Therefore, even this highly symmetric and minimal-size example provides results in a functionality that exceeds the standard framework that is limited to $\vec\chi_\mathrm{perm}$ and $\vec\chi_{\det}$.
    We systematically discuss the above linear decomposition more in the context of the next example.

\subsection{Asymmetric immanants in five modes}

    Next, we explore an example with two bosonic and three fermionic modes.
    This setting also yields contributions in the outcome probability distribution that are neither determinants nor permanents.
    The unitary network matrix in Eq. \eqref{eq:unitaryfb} then reduces to a $5\times5$ matrix $U^{\prime\dagger}$,
    \begin{equation}
        \label{eq:exampleumap}
        \begin{bmatrix}
    		\hat b_1^{\dagger}
    		\\
            \hat b_2^{\dagger}
            \\
    		\hat f_1^{\dagger}
    		\\
            \hat f_2^{\dagger}
    		\\
            \hat f_3^{\dagger}
    	\end{bmatrix}
        \mapsto
        \underbrace{
       \begin{bmatrix}
            \alpha_{11} & \alpha_{12} &\delta_{11} &\delta_{12} &\delta_{13}
            \\
            \alpha_{21} &\alpha_{22} &\delta_{21} &\delta_{22} &\delta_{23}
            \\
            \gamma_{11} &\gamma_{12} & \beta_{11} &\beta_{12} &\beta_{13}
            \\
            \gamma_{21} &\gamma_{22} &\beta_{21} & \beta_{22} & \beta_{23}
            \\
            \gamma_{31} & \gamma_{32} &\beta_{31} &\beta_{32} & \beta_{33}
       \end{bmatrix}
       }_{=U'^\dagger}
            \begin{bmatrix}
             \hat b_1^{\dagger}
            \\
            \hat b_2^{\dagger}
            \\
            \hat f_{1}^{\dagger}
    		\\
    		\hat f_2^{\dagger}

            \\
            \hat f_{3}^{\dagger}
       \end{bmatrix}.
    \end{equation}

    The input state considered is $|\Psi\rangle=\hat b_1^{\dag}\hat f_1^{\dag}\hat f_2^{\dag}|\mathrm{vac}\rangle=|1,0\rangle_{B}\otimes|1,1,0\rangle{F}=|1,0,1,1,0\rangle$, with a total excitation number of $N=3$, one excitation in a bosonic mode and two in fermionic modes.
    Then, the output state reads
    \begin{equation}
     \begin{aligned}
        |\Psi'\rangle
        =&
        (\alpha_{11}\hat{b}_1^{\dag}+\alpha_{12} \hat{b}_2^{\dag} +\delta_{11}\hat{f}_1^{\dag} +\delta_{12}\hat{f}_2^{\dag} +\delta_{13}\hat{f}_3^{\dag})&
        \\
        &\times(\gamma_{11}\hat{b}_1^{\dag}+\gamma_{12} \hat{b}_2^{\dag} +\beta_{11}\hat{f}_1^{\dag} +\beta_{12}\hat{f}_2^{\dag} +\beta_{13}\hat{f}_3^{\dag})&
        \\
        &\times(\gamma_{21}\hat{b}_1^{\dag}+\gamma_{22} \hat{b}_2^{\dag} +\beta_{21}\hat{f}_1^{\dag} +\beta_{22}\hat{f}_2^{\dag} +\beta_{23}\hat{f}_3^{\dag})|\mathrm{vac}\rangle,
     \end{aligned}
    \end{equation}
    using the unitary map in Eq. \eqref{eq:exampleumap}.

    As the measurement, we take the case of projections onto the output state $\langle \Pi|=\langle 10110|$.
    Then, the output probability $p=|\langle \Pi|\Psi'\rangle|^2$ can be computed and is determined by the overlap
    \begin{equation}
        \label{eq:probexample}
    \begin{aligned}
        {}&
        \langle \Pi|\Psi'\rangle
        \\
        ={}&
        \det
        \begin{bmatrix}
            \delta_{11} & \delta_{12} & \delta_{13}
            \\
            \beta_{11} & \beta_{12} & \beta_{13}
            \\
            \beta_{21} & \beta_{22} & \beta_{23}
        \end{bmatrix}
        +
        \mathrm{perm}
        \begin{bmatrix}
            \alpha_{11} & \alpha_{12} & \delta_{12}
            \\
            \gamma_{11} & \gamma_{12} & \beta_{12}
            \\
            \gamma_{21} & \gamma_{22} & \beta_{22}
        \end{bmatrix}
        \\
        {}&
        +
        \mathrm{perm}
        \begin{bmatrix}
            \alpha_{11} & \alpha_{12} & \delta_{11}
            \\
            \gamma_{11} & \gamma_{12} & \beta_{11}
            \\
            \gamma_{21} & \gamma_{22} & \beta_{21}
        \end{bmatrix}
        +
        \mathrm{perm}
        \begin{bmatrix}
            \alpha_{11} & \alpha_{12} & \delta_{13}
            \\
            \gamma_{11} & \gamma_{12} & \beta_{13}
            \\
            \gamma_{21} & \gamma_{22} & \beta_{23}
        \end{bmatrix}
        \\
        {}&
        +
        \mathrm{imm}_{\mathrm{FB}}
        \begin{bmatrix}
            \alpha_{12} & \delta_{11} & \delta_{12}
            \\
            \gamma_{12} & \beta_{11} & \beta_{12}
            \\
            \gamma_{22} & \beta_{21} & \beta_{22}
        \end{bmatrix}
       +
        \mathrm{imm}_{\mathrm{FB}}
        \begin{bmatrix}
            \alpha_{12} & \delta_{11} & \delta_{13}
            \\
            \gamma_{12} & \beta_{11} & \beta_{13}
            \\
            \gamma_{22} & \beta_{21} & \beta_{23}
        \end{bmatrix}
        \\
        {}&
        +
        \mathrm{imm}_{\mathrm{FB}}
        \begin{bmatrix}
            \alpha_{12} & \delta_{12} & \delta_{13}
            \\
            \gamma_{12} & \beta_{12} & \beta_{13}
            \\
            \gamma_{22} & \beta_{22} & \beta_{23}
        \end{bmatrix}
        +
        \mathrm{imm}_{\mathrm{FB}}
        \begin{bmatrix}
            \alpha_{11} & \delta_{11} & \delta_{12}
            \\
            \gamma_{11} & \beta_{11} & \beta_{12}
            \\
            \gamma_{21} & \beta_{21} & \beta_{21}
        \end{bmatrix}
        \\
        {}&
        +
        \mathrm{imm}_{\mathrm{FB}}
        \begin{bmatrix}
            \alpha_{11} & \delta_{11} & \delta_{13}
            \\
            \gamma_{11} & \beta_{11} & \beta_{13}
            \\
            \gamma_{21} & \beta_{21} & \beta_{23}
        \end{bmatrix}
        +
        \mathrm{imm}_{\mathrm{FB}}
        \begin{bmatrix}
            \alpha_{11} & \delta_{12} & \delta_{13}
            \\
            \gamma_{11} & \beta_{12} & \beta_{13}
            \\
            \gamma_{21} & \beta_{22} & \beta_{23}
        \end{bmatrix}.
    \end{aligned}
    \end{equation}
    This explicitly decomposes the overlap into known permanents and determinants as well as the immanant
    \begin{equation}
        \label{eq:imm3x3}
    \begin{aligned}
        \mathrm{imm}_{\mathrm{FB}}(A)
        ={}&
        A_{11}A_{22}A_{33}
        -A_{11}A_{23}A_{32}
        +A_{12}A_{21}A_{33}
        \\
        {}&
        +A_{12}A_{23}A_{31}
        -A_{13}A_{21}A_{32}
        -A_{13}A_{22}A_{31},
    \end{aligned}
    \end{equation}
    mixing bosonic and fermionic signatures in the terms.

    The expression in Eq. \eqref{eq:probexample} consists of determinants, permanents, and the immanant matrix function $\mathrm{imm}_{\mathrm{FB}}$, cf. Eq. \eqref{eq:imm3x3}.
    The latter expression resembles the determinant of a $3\times3$ matrix with two signs exchanged, likewise a permanent with three signs altered.
    The signs of the elements do not follow a clear convention for cycles of permutations.
    As stated in Ref. \cite{DL40}, a $3\times3$ matrix has three general immanants defined by the so-called character of the symmetric group $S_3$, the determinant, the permanent, and the standard immanant in Eq. \eqref{eq:Std3Imm}.
    Hence, by definition, the quantity $\mathrm{imm}_{\mathrm{FB}}$ is, in fact, not a immanant defined through the character.

\begin{figure}
   \centering
	\includegraphics[width=1.05\columnwidth]{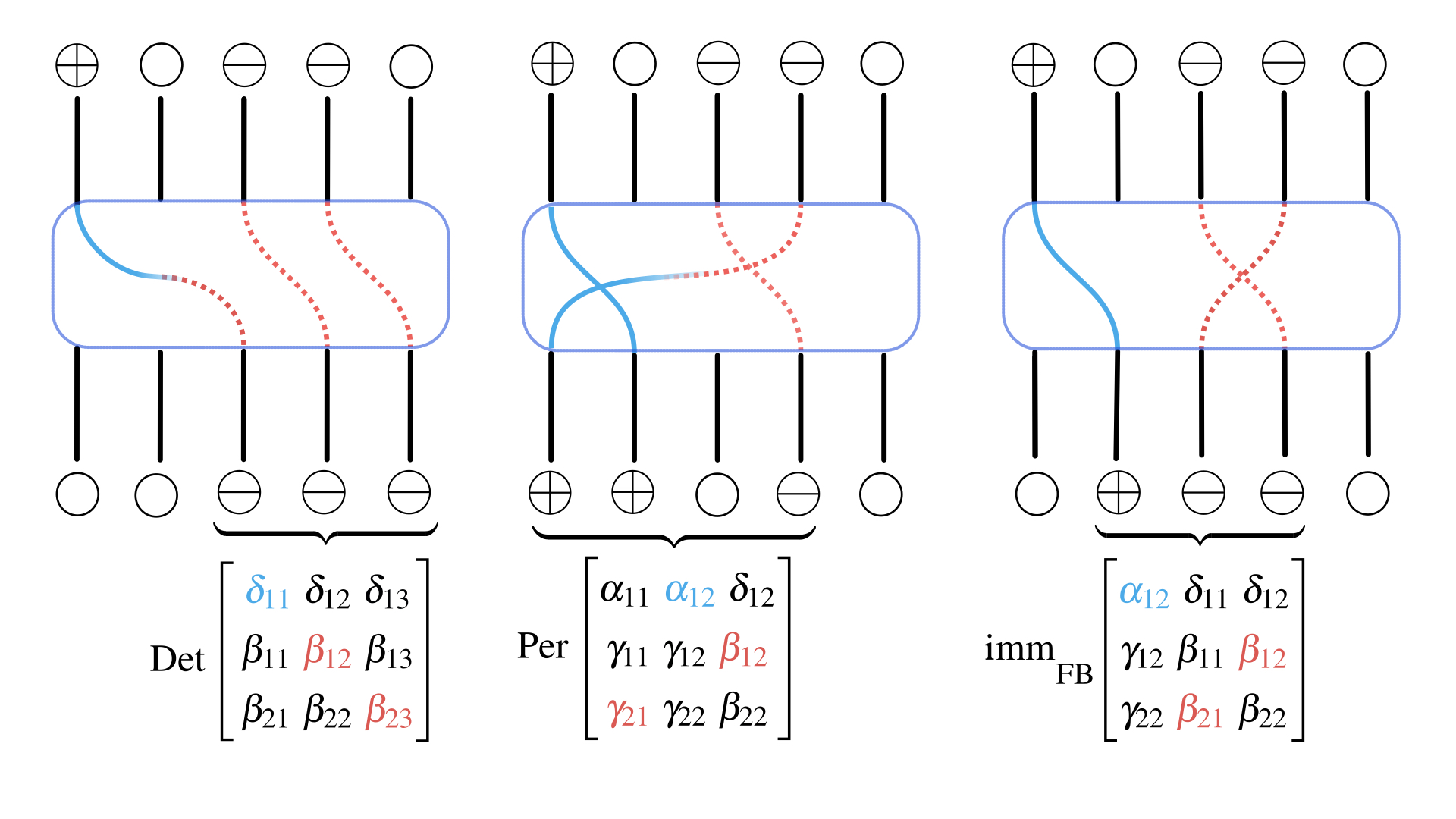}
	\caption{%
        Sketch of sampling network with two bosonic $\oplus$, three fermionic $\ominus$, and vacuum $\bigcirc$ modes drawn from top to bottom. 
        The three cases yield output probabilities for determinants (Det, left), permanents (Per, middle), and immanants (Imm, right), depending on the output-mode distribution.
        A purely fermionic output with one bosonic input transitioning to a fermionic mode yields the determinant.
        A mixed fermion boson output with two bosonic and one fermoinic excitation yields the permanent, whereas two fermions and one boson yield a general matrix function.
        The network paths are exemplary for the colored matrix element contributions, solid blue and dotted red lines referring to bosons and fermions, respectively. 
	}\label{fig:modes}
    \end{figure}

    Figure \ref{fig:modes} depicts which matrix functions contributes to which measured output presented in Eq. \eqref{eq:probexample}.
    There are special cases where a composite fermion boson network is only described by the determinant and permanent.
    Here, consistent with common fermion sampling, if the measurements only registers events at the fermionic output modes, only the determinant contributes to the output distribution. 
    The permanent, however, contributes to a mixed output of two bosons and one fermion since excitation operators of fermions and bosons commute, in addition to commuting bosonic operators.
    The same applies for the all-bosonic case.
    The immanant matrix functions can be found for cases with at least two fermions and at least one boson where the exchange symmetries create the non-trivial signs.
  
\subsection{Linear independence of matrix functions}

    A key component of the hybrid quantum sampling is that it is distinct from the interference of only bosons and only fermions.
    Therefore, we now quantify the linear independence from permanents and determinants and apply this to the previously introduced function $\mathrm{imm}_{\mathrm{FB}}$ [Eq. \eqref{eq:imm3x3}].

    Considering the expressions in Table \ref{tab:ThreeDimPermutations} as well as the vectors used in Eq. \eqref{eq:ThreeParticleDecomposition}, one straightforwardly observes that $\vec \chi_\mathrm{perm}$ and $\vec \chi_{\det}$ are orthogonal vectors.
    Thus, the contribution of a general matrix function free of permanent and determinant contributions, here dubbed $\mathrm{imm}_{\mathrm{free}}$, can be introduced as
    \begin{equation}
        \vec\chi_\mathrm{free}
        =
        \left(
            \mathbbm {1}
            -\frac{\vec \chi_\mathrm{perm}\vec \chi_\mathrm{perm}^\mathrm{T}}{\vec \chi_\mathrm{perm}^\mathrm{T}\vec \chi_\mathrm{perm}}
            -\frac{\vec \chi_{\det}\vec \chi_{\det}^\mathrm{T}}{\vec \chi_{\det}^\mathrm{T}\vec \chi_{\det}}
        \right)\vec\chi.
    \end{equation}
    This, for example, can be applied to $\vec\chi_\mathrm{FB}$, which defines the matrix function in Eq. \eqref{eq:imm3x3}, resulting in
    \begin{equation}
        \label{eq:freeBFterm}
        \vec\chi_\mathrm{free,FB}
        \propto
        \begin{bmatrix}
            1 & 2 & -1 & -1 & 1 & -2
        \end{bmatrix}^\mathrm{T}.
    \end{equation}
    Here, it also becomes evident that this vector is linearly independent from the standard immanant, $\vec\chi_{\text{Eq. \eqref{eq:Std3Imm}}}=\left[\begin{smallmatrix} 2 & 0 & 0 & 0 & -1 & -1 \end{smallmatrix}\right]^\mathrm{T}$, which itself is perpendicular to $\vec \chi_\mathrm{perm}$ and $\vec\chi_{\det}$.
    This additional independence is captured by the residual component,
    \begin{equation}
    \begin{aligned}
        {}&
        \vec \chi_\mathrm{residuum}
        \\
        ={}&
        \left(
            \mathbbm {1}
            -\frac{\vec \chi_\mathrm{perm}\vec \chi_\mathrm{perm}^\mathrm{T}}{\vec \chi_\mathrm{perm}^\mathrm{T}\vec \chi_\mathrm{perm}}
            -\frac{\vec \chi_{\det}\vec \chi_{\det}^\mathrm{T}}{\vec \chi_{\det}^\mathrm{T}\vec \chi_{\det}}
            -\frac{\vec\chi_{\text{Eq. \eqref{eq:Std3Imm}}}\vec\chi_{\text{Eq. \eqref{eq:Std3Imm}}}^\mathrm{T}}{\vec\chi_{\text{Eq. \eqref{eq:Std3Imm}}}^\mathrm{T}\vec\chi_{\text{Eq. \eqref{eq:Std3Imm}}}}
        \right)\vec\chi,
    \end{aligned}
    \end{equation}
    which we also used for the example shown in Eq. \eqref{eq:ThreeParticleDecomposition}.

\begin{figure}
	\includegraphics[width=1\columnwidth]{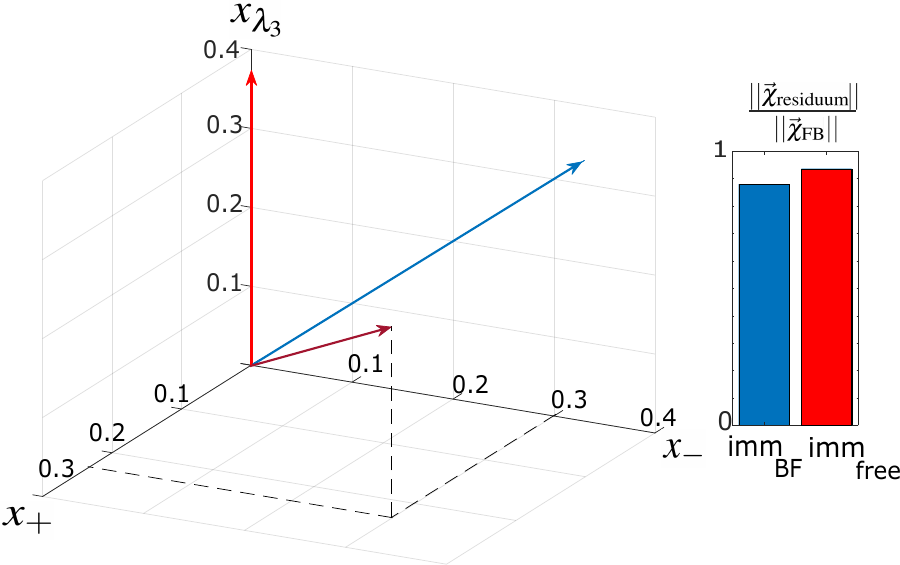}
	\caption{%
         Plot of the coefficients $x_+$ (zero), $x_-$ (blue arrow), and $x_{\lambda_3}$ (red arrow), jointly forming the purple vector, which is the expression in Eq. \eqref{eq:Fres} without the residual component.
         The bar plot features the permanent-and-determinant-free part (left) and residuum normalized to the full length of $\vec\chi_\mathrm{FB}$.
	}\label{fig:barplot}
    \end{figure}

    Through the aforementioned considerations, we can orthogonally decompose a the matrix function $\mathrm{imm}_\mathrm{FB}$ as
    \begin{equation}
        \label{eq:Fres}
        \vec \chi_\mathrm{FB}
        =
        x_+\vec \chi_\mathrm{perm}
        +x_-\vec \chi_{\det}
        +x_{\lambda_3}\vec\chi_{\text{Eq. \eqref{eq:Std3Imm}}}
        +\vec \chi_\mathrm{residuum},
    \end{equation}
    with expansion coefficients of the form $x_\nu=\vec\chi_\nu^\mathrm{T}\vec \chi_\mathrm{FB}/\vec\chi_\nu^\mathrm{T}\vec\chi_\nu$.
    Therein, the residual $\vec \chi_\mathrm{residuum}$ is a measure for the amount of contributions by the other remaining projected immanants of the $3!$-dimensional space with basis elements of permutations.
    In Fig. \ref{fig:barplot}, we visualize the components of $\chi_\mathrm{FB}$ in terms of determinant, permanent, and immanant contributions.
    This includes the free and non-zero residual parts from Eqs. \eqref{eq:freeBFterm} and \eqref{eq:Fres}.
    From this analysis, we conclude that the example function $\mathrm{imm}_\mathrm{FB}$ is composed of the determinant and standard immanant in Eq. \eqref{eq:Std3Imm} but does not include the permanent, with a relative residuum of $\|\vec\chi_\mathrm{residuum}||/\|\vec \chi_\mathrm{FB}\|=0.88$.

    In general, the determinant and permanent are upper and lower bounds for the computational complexity of immanants.
    The complexity immannants can depend on the irreducible character $\chi_\lambda$ and consequently whether it is more similar to the determinant or permanent \cite{PB00,SMCM13,NRA13,WH85,RC21}, which extends to functions $\chi: S_N\to \mathbb C$.
    That is, we quantify this through the overlap of $\vec \chi$ with the corresponding vectors of the permanent and determinant (i.e., the coefficients $x_\pm$).
    With respect to the other coefficient and the residuum, we currently cannot make specific statements regarding the complexity as they have not been studied in the literature to our knowledge.


\section{Towards quasiparticle sampling}
\label{sec:Quasiparticles}

    As a last generalization of quantum sampling problems in hybrid systems, we study quasiparticles which are defined as excitations of fields consisting of a superposition of bosons and fermions.
    This contrasts the previously investigated scenario of either purely bosonic or fermionic quantum statistics for each mode.
    For the sake of simplicity, we here consider the smallest, non-trivial example of $N=2$ excitations in $M=2$ modes, where $F=B=1$, to outline the unique quantum properties of quasiparticle sampling.

\paragraph{Physical motivation.}

    As a proof-of-concept example, we consider the seminal Jaynes--Cummings model \cite{JC63}, describing the linear interaction between one optical mode and one electronic two-level system.
    More specifically, the Jaynes--Cummings Hamiltonian can be diagonalized as
    \begin{equation}
    \begin{aligned}
        \hat H
        ={}&
        \hbar\omega_F \hat f^\dag \hat f
        +
        \hbar\omega_B \hat b^\dag \hat b
        +
        \hbar\kappa \hat f^\dag\hat b
        +
        \hbar\kappa^\ast \hat b^\dag\hat f
        \\
        ={}&
        \hbar\omega_1\hat a_1^\dag\hat a_1
        +
        \hbar\omega_2\hat a_2^\dag\hat a_2,
    \end{aligned}
    \end{equation}
    with frequencies $\omega_F=(\omega_1+|\lambda|^2\omega_2)/(1+|\lambda|^2)$ and $\omega_B=(|\lambda|^2\omega_1+\omega_2)/(1+|\lambda|^2)$ and a coupling constant $\kappa=\lambda(\omega_2-\omega_1)/(1+|\lambda|^2)$, where $\lambda\in\mathbb C\setminus\{0\}$.
    Importantly, the quasiparticle modes $\hat a_1$ and $\hat a_2$ are related to the fermionic mode $\hat f$ and the bosonic mode $\hat b$ via a unitary $V$
    \begin{equation}
        \begin{bmatrix}
            \hat a_1 \\ \hat a_2
        \end{bmatrix}
        =
        \underbrace{
        \frac{1}{\sqrt{1+|\lambda|^2}}
        \begin{bmatrix}
            1 & \lambda \\ -\lambda^\ast & 1
        \end{bmatrix}
        }_{=V}
        \begin{bmatrix}
            \hat f \\ \hat b
        \end{bmatrix}.
    \end{equation}
    Thus, neither quasiparticle mode can exhibit a purely fermionic or bosonic behavior.
    This has significant implications for quantum effects, such as entanglement \cite{BASS25}.
    Here, we are going to find that quantum sampling is affected in a sophisticated manner, too.

    For our following studies of sampling problems, we can explicitly compute the two-quasiparticle excitations
    \begin{equation}
        \label{eq:TwoQuasiparticles}
    \begin{aligned}
        \hat a_1^\dag\hat a_1^\dag|\mathrm{vac}\rangle
        ={}&
        \frac{
            \sqrt{2}\lambda^{\ast 2}|0,2\rangle
            +2\lambda|1,1\rangle
        }{1+|\lambda|^2},
        \\
        \hat a_2^\dag\hat a_2^\dag|\mathrm{vac}\rangle
        ={}&
        \frac{
            \sqrt{2}|0,2\rangle
            -2\lambda|1,1\rangle
        }{1+|\lambda|^2},
        \\
        \text{and}\quad
        \hat a_1^\dag\hat a_2^\dag|\mathrm{vac}\rangle
        ={}&
        \hat a_2^\dag\hat a_1^\dag|\mathrm{vac}\rangle
        \\
        ={}&
        \frac{
            \sqrt{2}\lambda^{\ast}|0,2\rangle
            +(1-|\lambda|^2)|1,1\rangle
        }{1+|\lambda|^2}.
    \end{aligned}
    \end{equation}
    Therein, we used Pauli's exclusion principle, $\hat f^{\dag 2}|\mathrm{vac}\rangle=0$.
    Note that the three vectors are spanned by only two vectors, $\sqrt2|0,2\rangle=\hat b^{\dag2}|\mathrm{vac}\rangle$ and $|1,1\rangle=\hat f^\dag\hat b^\dag|\mathrm{vac}\rangle=\hat b^\dag\hat f^\dag|\mathrm{vac}\rangle$, thus being linearly dependent, which alters the quantum statistics and behavior in the quasiparticle model \cite{BASS25}.

\paragraph{Quasiparticle sampling.}

    As done before, the quasiparticle modes as motivated above are propagated in a network,
    \begin{equation}
        \begin{bmatrix}
            \hat a_1^\dag \\ \hat a_2
        \end{bmatrix}
        \mapsto
        \begin{bmatrix}
            U_{1,1} & U_{1,2}
            \\
            U_{2,1} & U_{2,2}
        \end{bmatrix}
        \begin{bmatrix}
            \hat a_1^\dag \\ \hat a_2^\dag
        \end{bmatrix}.
    \end{equation}
    The input state consists of a single excitation in each mode, $|\Psi\rangle=\hat a_1^\dag\hat a_2^\dag|\mathrm{vac}\rangle$.
    Therefore, the output takes the form
    \begin{equation}
    \begin{aligned}
        |\Psi'\rangle
        ={}&
        (U_{1,1}\hat a_1^\dag + U_{1,2}\hat a_2^\dag)
        (U_{2,1}\hat a_1^\dag + U_{2,2}\hat a_2^\dag)|\mathrm{vac}\rangle
        \\
        ={}&
        U_{1,1}U_{2,1}\hat a_1^{\dag2}|\mathrm{vac}\rangle
        +
        U_{1,2}U_{2,2}\hat a_2^{\dag2}|\mathrm{vac}\rangle
        \\
        {}&
        +\left(
            U_{1,1}U_{2,2}+U_{1,2}U_{2,1}
        \right)\hat a_1^\dag\hat a_2^\dag|\mathrm{vac}\rangle.
    \end{aligned}
    \end{equation}

    Again, we want to determine the probability $p=|\langle\Pi|\Psi'\rangle|^2$ to find exactly one quasiparticle per output mode.
    Thus, the measurement is described by $|\Pi\rangle=\hat a_1^\dag\hat a_2^\dag|\mathrm{vac}\rangle$.
    Using the explicit expansions in Eq. \eqref{eq:TwoQuasiparticles}, we find
    \begin{equation}
    \begin{aligned}
        p
        \propto{}&
        \left|
            2\lambda^\ast U_{1,1}U_{2,1}
            +2|\lambda|^2\lambda U_{1,2}U_{2,2}
        \right.
        \\
        {}&
        \left.
            +\left(
                1+|\lambda|^4
            \right)\left(
                U_{1,1}U_{2,2}+U_{1,2}U_{2,1}
            \right)
        \right|^2
        =|E(U)|,
    \end{aligned}
    \end{equation}
    ignoring the normalizing denominator $(1+|\lambda|^2)^2$.
    The expression $E(U)$ that defines $p$ can be recast as
    \begin{equation}
        \label{eq:QuasiparticleImmanant}
        E(U)
        =\sum_{\rho\in F_2} \chi(\rho) \prod_{i=1}^2 U_{i,\rho(i)},
    \end{equation}
    where $F_2$ denotes all functions $\rho$, with the two-element set $\{1,2\}$ being the domain and co-domain of $\rho$.

\paragraph{Beyond permutations and immanants.}

    Solely permutations $S_2\subset F_2$ are used above in immanants for mixed fermion-boson sampling.
    This restricts the functions $\rho\in F_2$ [Eq. \eqref{eq:QuasiparticleImmanant}] to bijections $\rho=\sigma\in S_2$ [Eq. \eqref{eq:immmfunction}].
    For the case of quasiparticle sampling considered here, we obtain two additional maps which are not bijections.
    This is due to the superposition of fermionic and bosonic modes that define quasiparticles.
    Specifically, we have the two extra functions, both being constant, $\rho(i)=1$ and $\rho(i)=2$, and neither can be interpreted as a permutation.
    Thus, $E$ in Eq. \eqref{eq:QuasiparticleImmanant} even exceeds the general forms of immanants in Eq. \eqref{eq:immmfunction}.
    For the functions that correspond to permutations, we have $\chi(\rho)=1+|\lambda|^4$ in Eq. \eqref{eq:QuasiparticleImmanant}.
    This is the same value for both permutations (the identity and the swap), thus producing a term proportional to the permanent.
    For the two constant functions used in $p$, we have $\chi(\rho)=2\lambda^\ast\neq0$ and $\chi(\rho)=2|\lambda|^2 \lambda\neq0$, which has no counterpart in immanants.

    The proof-of-concept example, where $N=M=2$, can be generalized to an arbitrary number of quasiparticle modes ($M>2$) and number of quasiparticles ($N>2$) that propagate and interfere in the network.
    The resulting expression for $p$ is again proportional to $|E(U)|^2$, with $E(U)=\sum_{\rho\in F_N}\chi(\rho)\prod_{i=1}^N A_{i,\rho(i)}$ and $A_{m,n}=U_{j_n,k_m}$, in analogy to our previous expansion.
    Note that $S_N$ includes $N!$ permutations while $F_N$ includes $N^N$ functions, where $N!/N^N\stackrel{N\to\infty}{\longrightarrow}0$, showing that the number of terms in $E$ grows faster than for permanents, determinants, and immanants.


\section{Conclusion}
\label{sec:Conclusion}

    In summary, we considered quantum samplers in hybrid light-matter systems.
    That is, input excitations of both bosonic and fermionic modes can interact and interfere as they propagate.
    We then found general expressions for the sought-after output probability after the network propagation, describing generalized immanants, neither being determinant, permanent, nor standard immanants.
    This approach is based on the pairwise exchange symmetry of the involved bosonic, fermionic, and even anyonic modes.
    Several examples demonstrated how the different types of bosonic, fermionic, and other symmetries result in the total interference of the full hybrid multi-mode and multi-particle system.
    We also put forward a method to quantify the contribution of permanents, determinants, and standard immanants to generalized matrix functions that describe the output probabilities to gauge their complexity.
    Lastly, we studied hybrid quantum sampling based on quasiparticles, which can comprise joint light-matter excitations.
    Here, we found that the output probabilities are described via matrix functions which even exceed generalized immanants.

    Therefore, our investigation shows that hybrid samplers offer functionalities enabled by parastatistics which are inaccessible through bosonic and fermionic exchange symmetries alone.
    This might have an impact on complexity analysis, which can use the physical systems considered here for extended characterizations of multilinear functions of complex-valued transition matrices.
    Those can, for example, describe sophisticated graphs and transport scenarios for practical applications.
    For instance, this can include multi-walker quantum walks in which the exchange of the walkers has a non-trivial impact, such as relevant for vehicle routing with heterogeneous fleets \cite{MSNSVW26}.
    Moreover, a fundamental understanding the quantum interference across physical platform with distinct exchange symmetries is paramount for devising light-matter quantum interfaces for future quantum technologies.


\begin{acknowledgments}
	The authors acknowledge funding through the Deutsche Forschungsgemeinschaft (DFG, German Research Foundation) via the Transregional Collaborative Research Center TRR 142 (Projects No. A04 and No. C10, Grant No. 231447078).
\end{acknowledgments}


\end{document}